\documentclass[aps,prd,twocolumn,a4paper,superscriptaddress,floatfix]{revtex4-1}
\usepackage{graphicx}
\begin{document}

\newcommand{\be}{\begin{equation}}
\newcommand{\ee}{\end{equation}}
\newcommand{\bq}{\begin{eqnarray}}
\newcommand{\eq}{\end{eqnarray}}
\newcommand{\bsq}{\begin{subequations}}
\newcommand{\esq}{\end{subequations}}
\newcommand{\bc}{\begin{center}}
\newcommand{\ec}{\end{center}}

\title{Models for Small-Scale Structure on Cosmic Strings: I. Mathematical Formalism}
\author{C. J. A. P. Martins}
\email[Electronic address: ]{Carlos.Martins@astro.up.pt}
\affiliation{Centro de Astrof\'{\i}sica, Universidade do Porto, Rua das Estrelas, 4150-762 Porto, Portugal}
\author{E. P. S. Shellard}
\email[Electronic address: ]{E.P.S.Shellard@damtp.cam.ac.uk}
\affiliation{Department of Applied Mathematics and Theoretical Physics, Centre for Mathematical Sciences,\\ University of Cambridge, Wilberforce Road, Cambridge CB3 0WA, United Kingdom}
\author{J. P. P. Vieira}
\email[Electronic address: ]{up200907236@fc.up.pt}
\affiliation{Centro de Astrof\'{\i}sica, Universidade do Porto, Rua das Estrelas, 4150-762 Porto, Portugal}
\affiliation{Faculdade de Ci\^encias, Universidade do Porto, Rua do Campo Alegre, 4150-007 Porto, Portugal}

\date{28 May 2014}
\begin{abstract}
We describe the formalism of a quantitative analytic model for the evolution of realistic wiggly (as opposed to Goto-Nambu) cosmic strings. The model is particularly suited for describing the evolution of small-scale structure on string networks. We discuss model solutions in the extreme limit where the wiggles make up a high fraction of the total energy of the string network (which physically corresponds to the tensionless limit) and also provide a brief discussion of the opposite (linear) limit where wiggles are a small fraction of the total energy. A companion paper will discuss the detailed modelling and scaling behavior of the small-scale wiggles in the general model, together with a basic comparison with numerical simulations.
\end{abstract}
\maketitle

\section{\label{sint}Introduction}

Vortex-lines or topological strings can appear in a wide range of physical contexts, ranging from cosmic strings in the early universe to vortex-lines in superfluid helium (for reviews see \cite{VSH,COND1,COND2,GEYER}). Gaining a quantitative understanding of their important effects represents a significant challenge because of their nonlinear nature and interactions and because of the complexity of evolving string networks. This is particularly topical given the recent availability of high-quality data which one may use to constrain these models, such as that of the Planck satellite \cite{Planckdefects}.

Considerable reliance, therefore, has been placed on numerical simulations but unfortunately these turn out to be technically difficult and very computationally costly \cite{BB,AS,FRAC,RSB,VVO,Stuckey,Blanco,Hiramatsu}. This provides strong motivation for alternative analytic approaches, essentially abandoning the detailed \textit{statistical physics} of the string network to concentrate on its \textit{thermodynamics}. For the case of the simplest (Goto-Nambu) string networks the velocity-dependent one-scale (VOS) model \cite{MS1,MS2,MS3,MS4} has been exhaustively studied, and its quantitative success has been demonstrated by direct comparison with numerical simulations \cite{FRAC,ABELIAN}. It allows one to describe the scaling laws and large-scale properties of string networks in both cosmological and condensed matter settings with a minimal number of free parameters. More elaborate approaches can certainly be adopted \cite{ACK,POLR}, though usually at a cost of a larger number of free phenomenological parameters and/or (arguably) loss of intuitive clarity.

Almost all the work done on cosmic strings so far \cite{VSH} was concerned with the structureless Goto-Nambu strings. However, it is well known that cosmologically realistic string networks are not quite of Goto-Nambu type. Numerical simulations of cosmic strings in expanding universes \cite{BB,AS,FRAC,RSB,VVO,Stuckey,Blanco,Hiramatsu} have established beyond doubt the existence of a significant amount of short-wavelength propagation modes (that are commonly called \textit{wiggles}) on the strings, on scales that can be several orders of magnitude smaller than the correlation length.

This small-scale structure can be optimally described through its fractal properties \cite{FRAC}. On large scales we expect strings to be Brownian ($d_{large}\sim2$), while on small enough scales strings are smooth and locally straight (thus having $d_{small}\sim1$). Between these two scales, one finds an intermediate fractal region that extends over several orders of magnitude. This fractal region evolves in time, spreading out between the initial correlation length and the horizon size in such a way that any given physical scale is always loosing power. We stress that it is still not clear under which conditions small-scale structure continues building up indefinitely or eventually reaches a scaling solution like the large-scale properties of the network, although some progress has been made by a number of authors \cite{ACK,POLR}. The present article is a first step towards shedding some light on this issue.

A simple way of describing this small-scale structure is to measure the effective energy density at the scale of the correlation length. Note that this corresponds to measuring the ratio of the total energy to the \textit{bare} energy of a given piece of string. In units of the usual Goto-Nambu mass per unit length, the previous generation of numerical simulations \cite{BB,AS} suggested that
\begin{equation}
\mu_{rad}\sim1.9\, ,\qquad \mu_{mat}\sim1.5\,,
\label{tildemuold}
\end{equation}
respectively in the radiation and matter eras, while more recent and higher-resolution ones \cite{FRAC} suggest
\begin{equation}
\mu\sim1.25\, 
\label{tildemunew}
\end{equation}
in both epochs. The discrepancy is partly due to the higher resolution of the latter simulations but also to the slightly different definitions of the correlation length scale in both cases.

Be that as it may, it is clear that a substantial amount of the energy density of the string network can be in the form of these small-scale wiggles. One interesting consequence of this fact is that the typical velocity on the scale of the correlation length (usually called the \textit{coherent} velocity) is much smaller than the \textit{microscopic} velocity. In fact numerical simulations typically find \cite{BB,AS,FRAC}
\begin{equation}
v_{mic}\sim0.6\, ,\qquad v_{coh}\sim0.2\, .
\label{tildeves}
\end{equation}
It goes without saying that these factors can have a dramatic impact in the astrophysical and cosmological consequences of these networks, and hence in corresponding observational bounds.

Due to the very limited number of degrees of freedom available, the Goto-Nambu model cannot account for this phenomenology, nor for the build-up of charge and current densities which are expected to occur in many realistic string-forming particle physics models. More general string models \cite{CARTERA,CARTERB} are extremely useful for this purpose, and the much larger amount of algebra required is generously compensated by the resulting physical phenomenology. Two cases where these models have been shown to be useful are wiggly cosmic strings \cite{WIG1,WIG2} and superconducting strings \cite{WITT}. A first discussion of VOS-type models for these was presented in \cite{PHD}, and in the case of superconducting strings some further analysis can be found in \cite{SUPERC1,SUPERC2}.

Here we will focus on wiggly strings, and extend previous work \cite{PROC} to obtain the mathematical formalism necessary for a generalised VOS model that explicitly accounts for the build-up of small-scale structures on the strings. Specifically, this is done by fleshing out the effects of the effective energy density at the correlation scale ($\mu$) and obtaining an evolution equation for it, which will be coupled to the other two dynamical equations of the VOS model, for the characteristic length scale and the root-mean squared (RMS) velocity. Having done so we will discuss simple applications of this model, leaving a full discussion of how it applies to realistic networks in the early universe to a companion paper.

\section{\label{sevo}Goto-Nambu string evolution}

The velocity-dependent one-scale (VOS) model provides the most convenient and reliable method by which to calculate the large-scale quantitative properties of a Goto-Nambu string network in cosmological and other contexts \cite{MSCOND,MS4}. It is widely used for making quantitative predictions of the potential observational implications of cosmic strings \cite{VSH}. Given its simplicity, it is remarkable how well the VOS model performs when tested against high resolution numerical simulations of string networks \cite{FRAC,ABELIAN}. More recently, it has also been shown that this can be extended, if one is careful enough, to describe string networks in cosmological scenarios with extra dimensions \cite{XDIMR,XDIMW}. Here we will briefly review its mathematical formalism, so as to make clear how it is extended in the wiggly case.

The first assumption in this approach is to \textit{localise} the string so that we can treat it as a one-dimensional line-like object. This is clearly a good assumption for gauged strings, such as magnetic flux lines, but may seem more questionable for strings possessing long-range interactions, such as global strings or superfluid vortex lines.  However, good agreement between the VOS model and simulations has been established in both `local' and `global' cases \cite{MS4}.  

The second step is to \textit{average} the microscopic string equations of motion to derive the key evolution equations for the average string velocity $v$ and correlation length $L$. This is a generalisation of Kibble's original one-scale model \cite{KIB,BMOD}, and has been described in detail elsewhere \cite{MS2,MS3}. Kibble's model describes string motion in terms of a single \textit{characteristic length scale}, denoted $L$. In particular, it is assumed that this length scale coincides with the string \textit{correlation length} $\xi$ and the string \textit{curvature radius} $R$. We stress that this is an approximation which can be tested numerically \cite{FRAC,ABELIAN}. By incorporating a variable RMS velocity $v$, the VOS model extends its validity into early regimes with frictional damping and across the important matter-radiation transition, thus giving a quantitative picture of the complete history of a cosmic string network.

\subsection{Microscopic evolution}

The motion of a cosmic string with worldsheet coordinates $\sigma^a$ and background space-time coordinates $x^\mu$ with a metric $g_{\mu\nu}$ is obtainable from a variational principle applied to the action
\begin{equation}
S=\mu_0\int \sqrt{-\gamma}\, d^2\sigma  \, ,
\label{gnaction}
\end{equation}
where $\mu_0$ is the string mass per unit length, which one should in general expect to be of the order of the square of the symmetry breaking scale. We will be discussing strings in a background space-time where the line element is
\begin{equation}
ds^2=a^2\left(d\tau^2-{\bf{dx}}^2\right)\, \label{newlel2}
\end{equation}
and we will choose the usual temporal transverse gauge, that is
\begin{equation}
\sigma^0=\tau\, , \qquad {\bf {\dot x}}\cdot{\bf x}'=0 \, ,
\label{newgauge2}
\end{equation}
with dots and primes respectively denoting derivatives with respect to the time-like and space-like coordinates on the world-sheet. Recall that the coordinate energy per unit $\sigma$ is defined as
\begin{equation}
\epsilon^2=\frac{{\bf x}'{}^2}{1-{\dot {\bf x}}^2}\, .\label{ceus}
\end{equation}
Then one can show that the microscopic string equations of motion are
\begin{equation}
{\ddot {\bf x}}+2\frac{{\dot a}}{a}{\dot {\bf x}}(1-{\dot {\bf x}}^2)=\frac{1}{\epsilon}\left(\frac{{\bf x}'}{\epsilon}\right)'\, \label{geqn1}
\end{equation}
and
\begin{equation}
{\dot\epsilon}+2\epsilon\frac{\dot a}{a}{\dot {\bf x}}^2=0\, .\label{gneq2}
\end{equation}
For simplicity we are neglecting the effect of friction due to particle scattering, since this is only significant at early times. A thorough discussion of these effects may be found in \cite{MS2,MS3}

\subsection{\label{oldav}Averaged evolution}

As has been hinted above, the averaged quantities we use to describe the string network are its energy $E$ and RMS velocity $v$ defined by
\begin{equation}
E=\mu a(\tau)\int\epsilon d\sigma\, , \qquad v^2=\frac{\int{\dot{\bf x}}^2\epsilon d\sigma}{\int\epsilon d\sigma}\,.
\label{eee}
\end{equation}
Any string network divides fairly neatly into two distinct populations, \textit{viz.} long (or `infinite') strings and small closed loops. In the following we will always be discussing the long strings, except where explicitly stated otherwise.

The long string network is a Brownian random walk on large scales and can be characterized by a correlation length $L$. Bearing in mind the above assumptions, this can be used to replace the energy $E= \rho V$ in long strings in our averaged description, that is,
\be
\rho_\infty \equiv {\mu \over L^2}\,.
\ee
A phenomenological term must then be included to account for the loss of energy from long strings by the production of loops, which are much smaller than $L$. A \textit{loop chopping efficiency} parameter $c$ is introduced to characterize this loop production as
\begin{equation}
\left(\frac{d\rho}{dt}\right)_{\rm to\ loops}= c v\frac{\rho}{L}
\, . \label{rtl}
\end{equation}
In this approximation, we would expect the loop parameter $c$ to remain constant irrespective of the cosmic regime, because it is multiplied by factors which determine the string network self-interaction rate.

From the microscopic string equations of motion, one can then average to derive the evolution equation for the correlation length $L$,
\begin{equation}
2\frac{dL}{dt}=2HL(1 + {v^2})+c v \, , \label{evl0}
\end{equation}
where $H$ is the Hubble parameter. The first term in (\ref{evl0}) is due to the stretching of the network by the Hubble expansion which is modulated by the redshifting of the string velocity, while the second is the loop production term.

One can also derive an evolution equation for the long string velocity with only a little more than Newton's second law
\begin{equation}
\frac{dv}{dt}=\left(1-{v^2}\right)\left[\frac{k(v)}{L}-2Hv\right]\, ,
\label{evv0}
\end{equation}
where $k$ is called the \textit{momentum parameter}. The first term is the acceleration due to the curvature of the strings and the second is the damping term from the Hubble expansion. Note that strictly speaking it is the curvature radius $R$ which appears in the denominator of the first term. In the present context we are identifying $R=L$, but one should keep this distinction in mind in more general situations \cite{ABELIAN}. The parameter $k$ is defined by
\begin{equation}
k(v)\equiv\frac{\langle(1-{\dot {\bf x}^2})({\dot {\bf x}}\cdot {\bf u})\rangle} {v(1-v^2)}\, ,
\label{klod}
\end{equation}
with ${\dot {\bf x}}$ the microscopic string velocity and ${\bf u}$ a unit vector parallel to the curvature radius vector. For most relativistic regimes relevant to cosmic strings it is sufficient to define it as follows:
\begin{equation}
k_{\rm r}(v) =\frac{2\sqrt{2}}{\pi}\;\frac{1-8v^6}{1+8v^6}\,,
\label{krel}
\end{equation}
while in the extreme friction-dominated case ($v\rightarrow0$), we have the nonrelativistic limit
\begin{equation}
k_0 = \frac{2\sqrt{2}}{\pi}\, .
\label{knrel0}
\end{equation}
A detailed discussion of the generic form of the momentum parameter can be found in \cite{MS3}. 

\subsection{Linear scaling}

Scale-invariant solutions of the form $L\propto t$ (or $L\propto H^{-1}$) together with $v=const.$, only appear to exist when the scale factor is a power law of the form
\begin{equation}
a(t)\propto t^\lambda\, , \qquad 0<\lambda=const. <1\, \,.
\label{conda0}
\end{equation}
This condition implies that
\begin{equation}
L\propto t\propto H^{-1}\, ,
\label{props}
\end{equation}
with the proportionality factors dependent on $\lambda$. It is useful to introduce the following useful parameters to describe the relative correlation length and densities, defining them respectively as
\begin{equation}
L=\gamma t\,,  \qquad \zeta \equiv \frac{1}{\gamma^2}= \frac{\rho t^2}{\mu}\,.
\end{equation}
By looking for stable fixed points in the VOS equations, we can express the actual scaling solutions in the following implicit form:
\begin{equation}
\gamma^2_{GN}=\frac{k(k+c)}{4\lambda(1-\lambda)}\, ,\qquad v^2_{GN}=\frac{k(1-\lambda)}{\lambda(k+c)}\, ,
\label{scalsol}
\end{equation}
where $k$ is the constant value of $k(v)$ given by solving the second (implicit) equation for the velocity. It's easy to verify numerically that this solution is well-behaved and stable for all realistic parameter values. Finally, in such a scaling regime the string netork will be a fraction
\begin{equation}
\frac{\rho_{GN}}{\rho_{\rm crit}}= \frac{32\pi}{3k(k+c)}\frac{1-\lambda}{\lambda}\, G\mu
\end{equation}
of the universe's total energy  density.

If the scale factor is not a power law, then simple scale-invariant solutions like (\ref{scalsol}) do not exist. Physically this happens because the network dynamics are unable to adapt rapidly enough to the changes in the background cosmology. Examples of this are the transition between the radiation and matter-dominated eras and the onset of dark energy domination around the present day. Indeed, since this relaxation to a changing expansion rate is rather slow, realistic cosmic string networks are \textit{never} in scaling during the matter-dominated era \cite{MS3,FRAC}.

\section{\label{wigmic}Wiggly Cosmic Strings}

We can now re-visit the above discussion in the context of more general models. Further details can be found in \cite{CARTERA,CARTERB,PHD}. This section will concentrate on the microscopic dynamics of these models. The averaged evolution is considered in subsequent sections.

\subsection{Generalized Lagrangians\label{eldef}}

The motion of a cosmic string is in general obtainable from a variational principle applied to the action
\begin{equation}
S=-\int {\cal L}\sqrt{-\gamma}\, d^2\sigma  \, ;
\label{elaction}
\end{equation}
where the worldsheet metric is given by
\begin{equation}
\gamma_{ab}=g_{\mu\nu}x^\mu_{,a}x^\nu_{,b}  \, .
\label{elawsm}
\end{equation}
Quite generically \cite{CARTERA,CARTERB}, string models can be described by a Lagrangian density ${\cal L}$ depending only on the space-time metric $g_{\mu\nu}$, background fields such as a Maxwell-type gauge potential $A_\mu$ or a Kalb-Ramond gauge field $B_{\mu\nu}$ (but not their gradients) and any relevant internal fields, contained in a function $\Lambda$ (see below), that is
\begin{equation}
{\cal L}=\Lambda+J^\mu A_\mu+\frac{1}{2}W^{\mu\nu}B_{\mu\nu}+\ldots  \, .
\label{choiceon}
\end{equation}
The Maxwell and Kalb-Ramond fields are ideal for describing superconducting and global strings respectively, but it turns out that for our purposes they can be set to zero---the effect of small-scale structures can be encoded in the function $\Lambda$.

The simplest Goto-Nambu string model obviously corresponds to a constant Lagrangian density,
\begin{equation}
{\cal L}_{GN}=-\mu_0  \,.
\label{lagrgnamb}
\end{equation}
Models having a variable Lagrangian density are usually called elastic string models \cite{CARTERA,CARTERB}. The reason for this is that the energy density in the locally preferred string rest frame, which will henceforth be denoted by $U$, and the local string tension, denoted $T$, are constant for a Goto-Nambu string,
\begin{equation}
U=T=\mu_0  \, ,
\label{utgotonamb}
\end{equation}
but they are variable in general. In particular, one should expect that the string tension in an elastic model will be reduced with respect to the Goto-Nambu case due to the mechanical effect of the current.

Since elastic string models necessarily possess conserved currents \cite{CARTERA,CARTERB}, it is convenient to define a `stream function' $\phi$ on the world-sheet that will be constant along the current's flow lines. The part of the Lagrangian density ${\cal L}$ containing the internal fields is usually called the `master function', and can be defined as a function of the magnitude of the gradient of this stream function, $\Lambda=\Lambda(\chi)$, such that
\begin{equation}
\chi=\gamma^{ab}\phi_{,a} \phi_{,b}  \, ;
\label{eldefchi}
\end{equation}
notice that in more general cases with non-zero external fields these would be covariant derivatives. In our case, we will require a single scalar field, and the associated current can be pictured as a mass current. This means that \textit{we should think of wiggly strings as carrying a mass current}, which will renormalize the bare mass per unit length $\mu_0$. Indeed, the model with Lagrangian density
\begin{equation}
{\cal L}=-\mu_0\sqrt{1-\chi} \, , \label{translad}
\end{equation}
has the equation of state
\begin{equation}
UT=\mu_0^2 \, . \label{trseqnst2}
\end{equation}
and it has been shown that this equation of state arises in an \textit{exact} way \cite{MARTIN} in a macroscopic (in the sense of \textit{smoothed}) model of a wiggly string, that is a Goto-Nambu string containing a spectrum of small oscillations that one cannot (or does not want to) describe in microscopic detail \cite{WIG1,WIG2}.

Consistently with this physical picture, we will make the simplifying assumption that the potential $\chi$ depends only on the world-sheet time. We are therefore interpreting it as an \textit{effective} or renormalized quantity, defined on a scale that is smaller than the horizon (which is the scale beyond which the network is Brownian \cite{FRAC}) but still large enough (just) for the (small-scale) dependence on the space-like world-sheet coordinate to be negligible. In this phenomenological (and admittedly somewhat simplistic) sense it can be pictured as a \textit{mesoscopic} quantity. Possible ways to go beyond this approximation will be discussed in a subsequent publication.

\subsection{Microscopic evolution}

The free string equations of motion can now be obtained in the usual (variational) way. We retain the line element and gauge choice as in the previous section---see Eqns. (\ref{newlel2}--\ref{newgauge2})---and the coordinate energy per unit length along the string is still given by
\begin{equation}
\epsilon^2=\frac{{\bf x}'{}^2}{1-{\dot {\bf x}}^2}\, .\label{ceus1}
\end{equation}
The only difference (apart from the additional amount of algebra) is that there is now a further equation for the scalar field $\phi$. Indeed, rather than working with this directly it turns out to be convenient to define the dimensionless parameter $w$ by
\begin{equation}
\Lambda=-\mu_0 w\,;\label{defomg}
\end{equation}
and then the local string tension and energy density are simply given by
\begin{equation}
\frac{T}{\mu_0}=w\, ,\qquad \frac{U}{\mu_0}=\frac{1}{w}\, ,\label{wigt}
\end{equation}
so that
\begin{equation}
\frac{T}{U}=w^2\, .\label{wigu}
\end{equation}
Incidentally, notice that the equation of state for these networks has the form
\begin{equation}
3\frac{p}{\rho}=\left(1+\frac{T}{U}\right)v^2-\frac{T}{U}\, .\label{neweos}
\end{equation}
Hence wiggly strings still behave as radiation ($p/\rho\sim1/3$) in the ultra-relativistic limit. On the other hand, in the non-relativistic limit one has
\begin{equation}
\left(\frac{p}{\rho}\right)_{nr}=-\frac{1}{3}\frac{T}{U}\ge-\frac{1}{3}\, \label{nreos}
\end{equation}
while in the tensionless limit ($T/U\to0$)
\begin{equation}
\left(\frac{p}{\rho}\right)_{nt}=\frac{1}{3}v^2\,; \label{nteos}
\end{equation}
in particular, tensionless non-relativistic wiggly strings behave as matter ($p/\rho\sim0$). It may be of interest to assess the possible role of such strings in the context of the dark matter problem, but this is beyond the scope of the present work.

This being said, one can show that the microscopic string equations of motion are
\begin{equation}
{\ddot {\bf x}}+{\dot {\bf x}}(1-{\dot {\bf x}}^2)\frac{{\dot a}}{a}(1+w^2)=
\frac{w^2}{\epsilon}\left(\frac{{\bf x}'}{\epsilon}\right)'\, ,\label{wigeq1}
\end{equation}
\begin{equation}
\left(\frac{\epsilon}{w}\right)^{{\bf \dot {}}}+\left(\frac{\epsilon}{w}\right)\frac{\dot a}{a}
\left[2w^2{\dot {\bf x}}^2+(1+{\dot {\bf x}}^2)(1-w^2)\right]=0\, ,\label{wigeq2}
\end{equation}
with the dimensionless parameter $w$ obeying
\begin{equation}
\frac{{\dot w}}{w}=(1-w^2)\left(\frac{{\dot a}}{a}+
\frac{{\bf x}'\cdot {\dot{\bf x}}'}{{\bf x}'{}^2}\right)\, .\label{wigeq3}
\end{equation}
Alternatively, one can substitute Eq. (\ref{wigeq3}) into Eq. (\ref{wigeq2}) to obtain
\begin{equation}
\frac{\dot\epsilon}{\epsilon}+\frac{\dot a}{a}{\dot {\bf x}}^2(1+w^2)=(1-w^2)\frac{{\bf x}'\cdot {\dot{\bf x}}'}{{\bf x}'{}^2}\, ,\label{wigeq2alt}
\end{equation}
It is trivial to check that in the Goto-Nambu limit, $w=1$, we recover the original Eqns. (\ref{geqn1}--\ref{gneq2}).

\subsection{\label{wigcons}Simple consequences}

Now, the total energy of a piece of string is
\begin{equation}
E=a\int\epsilon Ud\sigma=\mu_0 a\int\frac{\epsilon}{w}d\sigma\, ;\label{enertot}
\end{equation}
note that part of this is the bare energy that  can be ascribed to the string itself,
\begin{equation}
E_0=\mu_0 a\int\epsilon d\sigma\, ,\label{enerst}
\end{equation}
while the rest is in the small-scale wiggles.
\begin{equation}
E_w=\mu_0 a\int\frac{1-w}{w}\epsilon d\sigma\,.\label{enerwg}
\end{equation}
Each of these energies can in principle be used to yield a characteristic length scale for the string network. For example, the total length could be the length that a Goto-Nambu string with the same total energy would have, while the bare length measures the actual length. \textit{Inter alia}, this has the immediate consequence that we will need phenomenological terms describing how energy is changed between the bare string and the wiggles.

For example, in the case of the original (Goto-Nambu) VOS model long string intercommutings did not affect the evolution of the network and so we did not need to directly model them. However, if one wants a model that explicitly includes small-scale structure, then one must consider these intercommutings, since it is well-known that any inter-commuting will increase the number of kinks on the string network---and consequently, from this point of view, add energy to the wiggles.

From the point of view of an analytic model, the key consequence of the existence of more than one length scale is that we are no longer allowed to identify the three natural length scales we considered in the Goto-Nambu case, namely a characteristic (energy) length scale $L$, the string correlation length $\xi$ and the string curvature radius $R$. In other words, \textit{we can no longer have a one-scale model}.

Moreover, one also needs to rethink the way in which averages are defined. Specifically, when one is defining average quantities over the string network (say, the average RMS velocity), should the average be over the total energy
\begin{equation}
\langle {\dot{\bf x}}^2 \rangle=\frac{\int{\dot{\bf x}}^2U\epsilon d\sigma}{\int U\epsilon d\sigma}\, ,\label{sexta1}
\end{equation}
or just the energy in string
\begin{equation}
\langle {\dot{\bf x}}^2 \rangle_0=\frac{\int
{\dot{\bf x}}^2\epsilon d\sigma}{\int\epsilon d\sigma}\, ? \label{sexta2}
\end{equation}
In other words, should pieces of string that have larger mass currents be given more weight in the average? Given the discussion so far, it should be intuitively clear that the first definition is the natural one, although the opposite choice deserves further discussion. These two different averaging procedures can be applied to any other relevant quantity. For a generic quantity $Q$, the two averaging methods are related via
\begin{equation}
\langle Q \rangle=\frac{\langle QU \rangle_0}{\langle U \rangle_0}\,.\label{twoaverages}
\end{equation}

An averaged model for wiggly cosmic string evolution should contain three (rather that two) evolution equations. Apart from evolution equations for a length scale and velocity, there will be a third equation which describes the evolution of small-scale structure. This is reminiscent of the three-scale model \cite{ACK}, but actually there are two crucial differences.

First, in the three scale model all three evolution equations do in fact describe length scales, while in our case only one of them does so (although a second equation can dependently be converted into one that does). Second, in the three scale model there is no allowance for the evolution of the string velocities.

From a physical point of view, the natural way to include small-scale structure in this type of analytic model is through an evolution equation for the renormalized string mass per unit length $\mu$, defined in the obvious way
\begin{equation}
\mu=\frac{E}{E_0}\, .\label{davmut}
\end{equation}

Before moving on to discuss the averaging of these equations for string networks and the corresponding macroscopic behavior, let us point out that simple solutions of these equations, for highly symmetric string configurations, have been discussed in \cite{PHD}.

Here we will only highlight a single result, \textit{viz.} that for a wiggly circular loop in flat space-time, and hence to a good approximation for any loops in the early universe that are much smaller than the cosmological horizon, the RMS velocity $v_{av}^2=\langle{\dot r}^2\rangle$ (the $\langle\rangle$ brackets denote an average over an oscillation period) and the ratio of the string tension and energy density ${\mu}^{-2}=\langle w\rangle^2$, which is a measure of the loop wiggliness, are related by
\begin{equation}
1-2v_{av}^2=\left(1-\frac{1}{\mu^2}\right)^2\, .\label{lopvmu}
\end{equation}
Observe that the standard result $v_{av}^2={1/2}$ is recovered in the Goto-Nambu limit $\mu=1$, but in general the effect of the wiggles is to slow the loops down. For example, for $\mu\sim1.25$ suggested by recent high-resolution simulations \cite{FRAC} we have $v_{av}^2\sim0.44$, while for $\mu\sim1.9$ (typical of earlier radiation era simulations \cite{BB,AS}) we would have $v_{av}^2\sim0.24$.

On the other hand, in the tensionless limit ($T/U\to0$, which corresponds to $\mu\to\infty$) we have
\begin{equation}
v_{av}^2\propto\mu^{-2}\, .\label{vmutens}
\end{equation}
In other words, we expect that a network that builds up so much small-scale structure that it becomes effectively tensionless will asymptotically freeze ($v\sim0$). Notice, however that such behavior is different from what is commonly called frustration, since that corresponds to an equation of state $p/\rho=-1/3$ while as we saw above a frozen tensionless network behaves as matter ($p/\rho=0$).

\section{\label{wigav}Averaged Evolution}

The averaging procedure for the transonic elastic model is in principle identical to the one followed for the Goto-Nambu case \cite{MS1,MS2,PHD}, although the added complexity will manifest itself in several ways. We will proceed fairly quickly, referring the reader to the above references for a thorough discussion, although we will of course point out the key differences between the two cases as they appear.

In accordance with the analysis in the previous section, we will define averaged quantities attributing more weight to regions with more small-scale structure. Hence we take the average of a generic quantity $Q$ to be
\begin{equation}
\langle Q\rangle=\frac{\int Q\frac{\epsilon}{w}d\sigma}{\int \frac{\epsilon}{w}d\sigma}\, . \label{avy}
\end{equation}
In particular, we will deal with the average RMS string velocity, $v^2=\langle {\dot {\bf x}}^2\rangle$ and also with the renormalised string mass per unit length
\begin{equation}
\mu\equiv\frac{E}{E_0}=\langle w\rangle^{-1}=\langle w^{-1}\rangle_0\, . \label{vwwv}
\end{equation}

Strictly speaking, $\mu$ is a scale-dependent quantity, $\mu=\mu(\ell,t)$ \cite{FRAC}, but following up from our discussion of the behavior of the mass current at the end of Sect. \ref{eldef} the $\mu$ thus defined is to be understood as a quantity measured at a mesoscopic scale somewhat smaller than the horizon. Intuitively, an obvious choice will therefore be the coherence length in the analytic model itself (more on this below).

\subsection{Network dynamics}

By differentiating the above equations, one finds the corresponding averaged evolution equations. Specifically, the total length of a given piece of string---or the corresponding density---evolves according to
\begin{widetext}
\begin{equation}
\frac{\dot E}{E}=\frac{\dot\rho}{\rho}+3\frac{\dot a}{a}=\frac{\dot E_0}{E_0}+\frac{\dot \mu}{\mu}=
\left[\langle w^2\rangle-v^2-\langle w^2{\dot {\bf x}}^2\rangle\right]\frac{\dot a}{a}\, , \label{wig_1}
\end{equation}
and we can immediately see that, apart from the extra evolution equation that we will get for $\mu$, we will also need to write a large number of averages involving mixed products of powers of $w$ and ${\dot {\bf x}}^2$ factors in terms of the individual quantities, $\mu$ and $v^2$.

We can similarly obtain the following evolution equation for the energy density in string
\begin{equation}
\frac{\dot E_0}{E_0}=\frac{\dot\rho_0}{\rho_0}+3\frac{\dot a}{a}=\left[1-\mu\langle w(1+w^2){\dot {\bf x}}^2\rangle\right]\frac{\dot a}{a}-\frac{a\mu}{R}\langle w(1-w^2)({\dot{\bf x}}\cdot\hat{\bf u})\rangle\, , \label{wig_20}
\end{equation}
where $R$ is the string curvature radius. As one would expect the Hubble expansion essentially acts on the string length, not the total length. In other words, stretching has the effect of decreasing wiggliness, just as it decreases velocity. On the other hand, curvature tends to accelerate the strings, thereby decreasing the string energy and hence tending to increase wiggliness.

Obviously the evolution equation for $\mu$ is in principle not independent, and can be obtained from the above. However, one must be careful about the choice of lengthscale at which one is defining it. If this is a fixed scale, then all one has to do is use the definition (Eq. \ref{vwwv}) and take the difference of the dynamical equations for $E$ and $E_0$. \textit{However, if we want to define it at the scale of the correlation length we must allow for the fact that this scale also evolves with time.} It follows that generically we have
\begin{equation}
\frac{\dot\mu}{\mu}=\frac{\dot E}{E}-\frac{\dot E_0}{E_0}+\frac{1}{\mu}\frac{\partial\mu}{\partial\ell}{\dot\ell}\,, \label{wig_40}
\end{equation}
where $\ell$ is the mesoscopic lengthscale at which we're effectively defining $\mu$. The energy terms can now be obtained from the above equations, while the scale drift term can be related to the multifractal dimension (see for example \cite{TAKAYASU}), to yield, to second order in $\ell/R$,
\begin{equation}
\frac{\dot\mu}{\mu}=\frac{a\mu}{R}\langle w(1-w^2)({\dot{\bf x}}\cdot\hat{\bf u})\rangle+\frac{\dot a}{a}\left[\langle w^2\rangle-1+\langle (\mu w-1)(1+w^2){\dot {\bf x}}^2\rangle\right] + [d_m(\ell)-1]\frac{\dot\ell}{\ell}  \,. \label{wig_4}
\end{equation}
This also means that a drift term (with the opposite sign) should be included in the equation for the density in string
\begin{equation}
\frac{\dot E_0}{E_0}=\frac{\dot\rho_0}{\rho_0}+3\frac{\dot a}{a}=\left[1-\mu\langle w(1+w^2){\dot {\bf x}}^2\rangle\right]\frac{\dot a}{a}-\frac{a\mu}{R}\langle w(1-w^2)({\dot{\bf x}}\cdot\hat{\bf u})\rangle - [d_m(\ell)-1]\frac{\dot\ell}{\ell}     \, , \label{wig_2}
\end{equation}
Notice that this makes physical sense: $E$ is the total energy if the network, and thus an invariant quantity, but the string energy $E_0$ and $\mu$ do depend on the scale at which we have decided to measure them.

Finally the evolution equation for the string RMS velocity now has the form
\begin{equation}
\dot{\left(v^2\right)}=\frac{2a}{R}\langle w^2(1-{\dot {\bf x}}^2)({\dot{\bf x}}\cdot\hat{\bf u})\rangle -\frac{\dot a}{a}\langle (v^2+{\dot {\bf x}}^2)(1+w^2)(1-{\dot {\bf x}}^2)\rangle\, , \label{wig_3}
\end{equation}
\end{widetext}
and for analogous reasons there is also a scale drift term in this equation which has the form
\begin{equation}
\frac{\partial v^{2}}{\partial l}\frac{dl}{dt}=\frac{1-v^{2}}{1+\left\langle w^{2}\right\rangle }\frac{\partial\left\langle w^{2}\right\rangle }{\partial l}\frac{dl}{dt}\,. \label{v_drift}
\end{equation}
The physical interpretation of this term is not as simple as that for the renormalized mass, but one immediate consequence of the presence of this drift term is that strictly speaking this is no longer a purely 'microscopic' RMS velocity, but rather a 'mesoscopic' one.

The coefficient in the drift term is unity for a Brownian network ($d_m=2$) and vanishes for straight segments ($d_m=1$): a straight line is a straight line regardless of the scale at which one is looking at it. Naturally it also vanishes if we're considering a time-independent scale. The analogies between the evolution equations for $\mu$ and $v$ are manifest. Observe, however, an expected but crucial difference: the curvature term in the wiggliness equation vanishes both in the tensionless and the Goto-Nambu limits, while that in the velocity equation only vanishes in the tensionless limit.

Recall that $v$ and $\mu$ are averaged quantities; they have been put inside average signs, respectively in Eqns (\ref{wig_3}) and (\ref{wig_4}) simply as a means to yield simpler algebraic expressions; when expanding those expressions they can be freely taken out of the averages since they have no spatial dependence. Moreover, note that in order to obtain the terms involving the curvature radius $R$ in the above equations one needs to make use of the following identities
\begin{equation}
\frac{1}{\epsilon(1-{\dot {\bf x}}^2)}\left(\frac{{\bf x}'}{\epsilon}\right)'\cdot{\dot {\bf x}}=-\frac{{\bf x}'\cdot{\dot {\bf x}}'}{{\bf x}'{}^2}
=\frac{a}{R}({\dot{\bf x}}\cdot\hat{\bf u})\, , \label{wig_5}
\end{equation}
where $\hat{\bf u}$ is a unit vector defined as
\begin{equation}
\frac{a}{R}\hat{\bf u}=\frac{d^2{\bf x}}{ds^2}
\, \label{crvnewww}
\end{equation}
and
\begin{equation}
ds=|{\bf x}'|d\sigma=\sqrt{1-{\dot {\bf x}}^2}\epsilon d\sigma \,. \label{crvnewww2}
\end{equation}
These have been discussed in more detail in \cite{MS2,PHD}. 

\subsection{\label{enloss}Energy Loss Phenomenology}

We still need to discuss what phenomenological terms should be added to the above equations to account for energy losses into loops and for the energy transfer between the bare string and the wiggles. (Recall that we are focusing our analysis on the evolution of the long string network.) The discussion is now slightly more elaborate than in the Goto-Nambu case. Firstly we need to define the string correlation length, which clearly needs to be defined with respect to the  bare (as opposed to the total) string density. Specifically we have
\begin{equation}
\rho_0\equiv\frac{\mu_0}{\xi^2} \, . \label{wig_b2}
\end{equation}
We will then make the further assumption that the correlation length thus defined is approximately equal to the string curvature radius defined in Eqn. (\ref{crvnewww}) and appearing in Eqns. (\ref{wig_2}--\ref{wig_4}),
\begin{equation}
\xi\sim R \, ; \label{wig_b3}
\end{equation}
such an assumption can be tested numerically \cite{FRAC}, and although not exact is sufficiently accurate for our present purposes. (Nevertheless, we do stress that there are situations where it may break down dramatically---we shall discuss such contexts in future work.) The correlation length $\xi$ still has a physically clear meaning, while the other characteristic length scale $L$ is now only a proxy for the total energy in the network.

In analogy with what was done in Eqn. (\ref{rtl}) for the simple Goto-Nambu case, we define the fraction of the bare energy density lost into loops per unit time as
\begin{equation}
\left(\frac{1}{\rho_0}\frac{d\rho_0}{dt}\right)_{\rm loops}=-c f_0(\mu) \frac{v}{\xi}\, . \label{wig_b4}
\end{equation}
Numerical simulations suggest that small-scale structure enhances loop production, and we phenomenologically allow for this possible enhancement by allowing for an explicit dependence on $\mu$, encoded in a function $f_0(\mu)$ which should approach unity in the Goto-Nambu limit. The behavior of this function will be further discussed in the companion paper. It may also be possible to infer it from numerical simulations with sufficiently high resolution.

Importantly, in the wiggly case we have an additional phenomenological term. Whenever two strings inter-commute, kinks are produced (whether or not loop production occurs). From the point of view of our model, this corresponds to energy being transferred from the bare string to the small-scale wiggles. We will model this transfer from the bare string into the small-scale wiggles as follows
\begin{equation}
\left(\frac{1}{\rho_0}\frac{d\rho_0}{dt}\right)_{\rm wiggles}= -cs(\mu)\frac{v}{\xi}\, , \label{wig_b6}
\end{equation}
in analogy with the above term for losses into loops. Beyond the fact that the phenomenological parameter $s$ vanishes in the Goto-Nambu limit, its precise behavior is less obvious than the former one. Note that in particular it should include the effects of kink decay on long strings (notably due to gravitational radiation), a process that is not accounted for in numerical simulations of string networks. Further work will be required to understand this parameter more thoroughly. 

As for the fraction of the total energy lost into loops, we need to take into account that the energy may be come from the bare string or from the wiggles
\begin{equation}
\left(\frac{1}{\rho}\frac{d\rho}{dt}\right)_{\rm loops}=\left(\frac{1}{\rho}\frac{d\rho_0}{dt}\right)_{\rm loops}+\left(\frac{1}{\rho}\frac{d\rho_w}{dt}\right)_{\rm loops}\,. \label{wig_b5a}
\end{equation}
The energy loss from the bare string has already been characterized by the parameter $f_0$ in Eq. (\ref{wig_b4}); defining an analogous term for the losses from the wiggles
\begin{equation}
\left(\frac{1}{\rho_w}\frac{d\rho_w}{dt}\right)_{\rm loops}=-c f_1(\mu) \frac{v}{\xi}\,, \label{wig_b5b}
\end{equation}
we end up with
\begin{equation}
\left(\frac{1}{\rho}\frac{d\rho}{dt}\right)_{\rm loops}=-c\left[\frac{f_0}{\mu}+f_1\left(1-\frac{1}{\mu}\right)\right]\frac{v}{\xi}=-c f(\mu) \frac{v}{\xi}\,, \label{wig_b5}
\end{equation}
where we defined an overall loss parameter $f$. We might expect this term to have a stronger dependence on $\mu$ than that of $f_0$, to account for the fact that loops are preferentially produced from regions of the long string network containing more small-scale structure than average. There is clear evidence of this fact from numerical simulations \cite{BB,AS,FRAC}. Somewhat similar parameters have been introduced before \cite{ACK}; these are usually constant and defined as the excess kinkiness of a loop compared to a piece of long string of the same size. Here, we will explicitly include a $\mu$ dependence.

As a simple illustration, if we fix $f_0=1$, specify that both energy loss terms strictly match the Goto-Nambu case, and recall that
\begin{equation}
\xi^2=\mu L^2\,,
\end{equation}
we immediately get
\begin{equation}
f(\mu)=\sqrt{\mu}\,, \label{wig_b5gnb}
\end{equation}
and therefore
\begin{equation}
f_1(\mu)=\frac{\mu^{3/2}-1}{\mu-1}\,, \label{wig_b5gna}
\end{equation}
but again we emphasize that the detailed form of these functions warrants further discussion, and should be checked in high-resolution numerical simulations.

\section{\label{tless}The tensionless limit}

Before going into a full analysis of the wiggly model (which we leave for a companion paper), it is instructive to look into particular limits of the model. In this section we study the tensionless limit where most of the energy is in the small-scale wiggles ($w\to0$, $T/U\to0$), and in the following one what we call the linear limit. Both of these limits have the advantage of substantially simplifying the dynamical equations while at the same time allowing clear physical interpretations of the remaining terms and the corresponing solutions.

The limit $w\to0$ physically corresponds to the local string tension being negligible when compared to the energy density, $T/U\to0$, and hence to very high wiggliness, $\mu\gg1$. This is in fact a very simple limit to study (for example the scale drift term in the velocity equation is negligible), but it will provide insights into the behavior of wiggles that will be very useful for future studies.

We can start with the evolution equation for the RMS velocity (Eqn. \ref{wig_3}), which yields
\begin{equation}
v\propto a^{-1} \, ; \label{tensnl1}
\end{equation}
and using this result in the equation for the total energy (Eqn. \ref{wig_1}) then yields
\begin{equation}
E={\rm const.}\,,\quad \rho\propto a^{-3} \, . \label{tensnl2}
\end{equation}
What is happening is physically obvious. The network is conformally stretched, and since the stretching acts on the string length the wiggliness will in fact have to be decreasing. Moreover, since the network is frozen it will eventually dominate the energy density of the universe. Solving the Friedmann equation with the string density, we immediately find
\begin{equation}
a\propto t^{2/3} \, , \label{tensnlhubble}
\end{equation}
in other words, this string-dominated universe is like a matter-dominated universe. This is of course not a surprise: we have already seen above that the equation of state for wiggy strings in the tensionless limit is that of matter, $p/\rho\to0$ (cf. Eq. \ref{nteos}).

This physical interpretation can be confirmed by looking at the remaining equations. We start by looking at what happens on a fixed scale, and we easily find
\begin{equation}
E_0\propto a\,,\quad \rho_0\propto a^{-2} \, . \label{tensnl4}
\end{equation}
It then follows that the network's correlation length is proportional to the scale factor
\begin{equation}
\xi\propto a \, , \label{tensnl5}
\end{equation}
as indeed is the case for the total length in string
\begin{equation}
\ell_0=\frac{E_0}{\mu_0}=\frac{E}{U}\propto a \,, \label{tensnl6}
\end{equation}
which confirms the conformal stretching. As for the renormalized mass per unit length, since $\mu E_0=E={\rm const.}$ we immediately have in this fixed scale case
\begin{equation}
\mu\propto a^{-1} \, ; \label{tensnl3}
\end{equation}
recall the mesoscopic interpretation of $\mu$: since the network is frozen but being stretched by the expansion, the effective mass per unit length on a given scale is correspondingly reduced.

It's important to understand a key difference between this case and the one previously discussed at the end of Sect. \ref{wigcons}, specifically concerning Eqn. (\ref{lopvmu}). Recall that in the tensionless limit of that case we had found $v\propto \mu^{-1}$. That situation effectively corresponds to a Minkowski space limit (small scales, well below the horizon) and the effect of wiggliness is to reduce the oscillation period of otherwise free loops. In the present section we are effectively taking the opposite (super-horizon) limit. Consequently expansion is not only important but indeed crucial, and the string velocity and wiggliness (on a fixed scale) are both damped, leading to $\mu\propto v\propto a^{-1}$.

In the general case where the scale where $\mu$ is defined is allowed to vary, we have 
\begin{equation}
\frac{\dot E_0}{E_0}=-\frac{\dot\mu}{\mu}= \frac{\dot a}{a} - \left[d_m(\ell)-1\right]\frac{\dot\ell}{\ell}\,. \label{wig_xi}
\end{equation}
If we follow a scale that is proportional to the scale factor ($\ell\propto a$), we find the suggestive
\begin{equation}
\frac{\dot\mu}{\mu}=\left[d_m(\ell)-2\right]\frac{\dot a}{a}\,, \label{wig_x2}
\end{equation}
and for a Brownian network $\mu$ and $E_0$ would both be constant. On the other hand $\mu$ will decrease (increase) for a network with a smaller (larger) fractal dimension, and $E_0$ will have the opposite behaviour. Assuming for simplicity a constant multifractal dimension we can in fact write
\begin{equation}
\mu\propto a^{d_m-2}\,,\quad \rho_0\propto a^{-(1+d_m)}\,, \label{wig_x3}
\end{equation}
or equivalentely
\begin{equation}
\xi\propto a^{(1+d_m)/2}\,; \label{wig_x4}
\end{equation}
and therefore the characteristic lengthscale $L$ behaves as
\begin{equation}
L\propto a^{3/2}\propto t\,, \label{wig_x5}
\end{equation}
independently of the averaging scale and consistently with the aforementioned behaviour of the total energy $E$.

As an alternative, if we instead follow the horizon scale ($\ell\propto d_H$), for $a\propto t^{2/3}$ we have $d_H(t)=3t$, and relating this to the scale factor we find (still assuming a constant fractal dimension)
\begin{equation}
\mu\propto a^{(3d_m-5)/2}\,,\quad \xi\propto a^{(1+3d_m)/4}\,; \label{wig_x6}
\end{equation}
now $\mu$ will be a constant for a multifractal dimension $d_m=5/3$. Interestingly, this is the fractal dimension of a self-avoiding random walk in three dimensions \cite{TAKAYASU}. Conversely, if we follow the scale of the correlation length itself $\ell\propto\xi$ we find
\begin{equation}
\mu\propto a^{4/(3-d_m)-3}\,,\quad \xi\propto a^{2/(3-d_m)}\,, \label{wig_x7}
\end{equation}
and again a constant $\mu$ will correspond to a self-avoiding random walk, $d_m=5/3$.

\section{\label{linlim}The linear limit}

The most interesting limit is what we'll refer to as the linear limit ($w\to1$, $\mu\to1$) where the wiggliness is small and can be treated as a linear order perturbation to the Goto-Nambu case. In this case the energy in the wiggles is only a small fraction of the total energy, and can therefore be treated as a small perturbation of the Goto-Nambu case. This limit may also be reasonable as an approximation for comparisons with numerical simulations, which typically start out with very little or no small-scale wiggles.

We will therefore take the linear limit of our evolution equations in the wiggliness parameter, thereby obtaining a model that is much simpler than the generic case but may still contain most of the relevant physics. At the microscopic level, let's define
\begin{equation}
w=1-y \, , \label{linear1}
\end{equation}
where $y\ll1$; macroscopically this corresponds to
\begin{equation}
\mu\approx 1+\langle y\rangle\equiv 1+Y \, , \label{linear2}
\end{equation}
where $Y$ is similarly small and positive, and averaged quantities are now defined as
\begin{equation}
\langle Q\rangle=\frac{\int Q (1+y)\epsilon d\sigma}{\int (1+y) \epsilon d\sigma}\sim\langle Q \rangle_0+corr_0(y,Q)\, . \label{avylin}
\end{equation}
As expected, to first approximation the two previously discussed averaging procedures are now equivalent for quantities independent of $w$. This also implies that the cross-terms now become trivial if we assume ${\dot {\bf x}}^2$ to be independent of $w$
\begin{equation}
\langle w^{\alpha_1} {\dot {\bf x}}^{2\alpha_2}\rangle\sim\langle w^{\alpha_1} \rangle\langle {\dot {\bf x}}^{2\alpha_2}\rangle\sim (1-\alpha_1 Y)v^{2\alpha_2}\, , \label{newcross}
\end{equation}
since the neglected terms are all higher-order. As a simple example of the behavior in this linear limit, the relation between the wiggliness and the averaged velocity for a circular loop, given by Eqn (\ref{lopvmu}), becomes
\begin{equation}
v_{av}^2\sim\frac{1}{2}\left(1-Y^2\right)\,.\label{linlopvmu}
\end{equation}

In what follows we will be assuming that the velocity dependence of the parameter $k(v)$ remains unchanged from its standard form, which we discussed at the end of Sect. \ref{oldav}. Specifically the assumption is that $k$ does not depend on the wiggliness beyond the dependencies on $w$ that are explicit in the curvature terms in Eqns (\ref{wig_2}--\ref{wig_4}). This is an assumption that is adequate for our purposes in this paper, but one that needs to be improved upon in the future (particularly when dealing with cases of high wiggliness).

One can then proceed to linearize the averaged evolution equations (Eqns. \ref{wig_1}--\ref{wig_4}), and one finds after some straightforward algebra
\begin{equation}
\frac{\dot E}{E}=\left[(1-2v^2)-2Y(1-v^2)\right]\frac{\dot a}{a}\, \label{wig_1lin}
\end{equation}
\begin{equation}
\frac{\dot E_0}{E_0}=\left[(1-2v^2)+2Yv^2\right]\frac{\dot a}{a}-2\frac{kaYv}{R} -\left[d_m(\ell)-1\right]\frac{\dot\ell}{\ell}   \, , \label{wig_2lin}
\end{equation}
\begin{widetext}
\begin{equation}
\dot{\left(v^2\right)}=2v(1-v^2)\left[\frac{ka}{R}(1-2Y)-2v(1-Y)\frac{\dot a}{a}-\frac{1+2Y}{2v}\left[d_m(\ell)-1\right]\frac{\dot\ell}{\ell}\right]\, \label{wig_3lin}
\end{equation}
\end{widetext}
\begin{equation}
{\dot Y}=2Y\left(\frac{kav}{R}-\frac{\dot a}{a}\right) +\left[d_m(\ell)-1\right]\frac{\dot\ell}{\ell} \,; \label{wig_4lin}
\end{equation}
one can trivially check that the standard dynamical equations are recovered in the Goto-Nambu limit, $Y=0$. (Note that in this case also $d_m=1$.)

Finally, switching from conformal to physical time and introducing the energy loss terms in accordance with the discussion of Sect. \ref{enloss} we end up with the generalized (linear wiggly) VOS model evolution equations
\begin{widetext}
\begin{equation}
2\frac{dL}{dt}=2[1+v^2+Y(1-v^2)]HL+cfv\left(1-\frac{1}{2}Y\right)\, \label{newvos0}
\end{equation}
\begin{equation}
2\frac{d\xi}{dt}=2[1+(1-Y)v^2]H\xi+[2kY+c(f_0+s)]v+\xi \left[d_m(\ell)-1\right]\frac{1}{\ell} \frac{d\ell}{dt}  \, \label{newvos1}
\end{equation}
\begin{equation}
\frac{dv}{dt}=(1-v^2)\left[\frac{k}{\xi}(1-2Y)-2Hv(1-Y)-(d_m(\ell)-1)\frac{1+2Y}{2v\ell}\frac{d\ell}{dt} \right] \,, \label{newvos2}
\end{equation}
\begin{equation}
\frac{dY}{dt}=[2kY+c(f_0+s-f)]\frac{v}{\xi}-2HY + \left[d_m(\ell)-1\right]\frac{1}{\ell} \frac{d\ell}{dt}    \, , \label{newvos3}
\end{equation}
\end{widetext}
which again reduce to the original VOS model equations for $Y=0$. Recall that $f_0$, $f$ and $s$ are in principle functions of $Y$ (cf. the discussion of of Sect. \ref{enloss}) while $k$ is a function of velocity. Note that we expect both $(f_0+s-f)$ and $(d_m-1)$ to be linear in $Y$.

A simple analysis of the scaling behavior of these evolution
equations can now be carried out. It makes sense to consider a situation
in which $\ell\propto t$ and look for the solution with $\xi=\gamma_{\xi}t$,
$v=const$, and $Y=const$. In order to solve this we assume $d_{m}\sim1+2Y$,
which numerical simulations \cite{FRAC} suggest may be a reasonable approximation.
If $\gamma_{GN}$ and $v_{GN}$ are the scaling parameters in the Goto-Nambu
case (refer to Eqn. (\ref{scalsol}) for their definitions), then to first order in $Y$ the new (wiggly) scaling parameters
are obtained (for $c\neq0$) by solving the algebraic system
\begin{equation}
\gamma_{\xi}\sim\frac{cfv}{2\left[1-\lambda\left(1+v^{2}+\left(1-v^{2}\right)Y\right)\right]}\label{eq:gam_corr}
\end{equation}
\begin{equation}
\frac{v^{2}}{v_{GN}^{2}}\sim1+\left[\frac{c\lambda\left(2-A-2B\right)-2k\left(1-2\lambda\right)}{2\lambda\left(k+c\right)}\right]Y\label{eq:v_corr}
\end{equation}
\begin{widetext}
\begin{equation}
Y\sim\frac{2\left(\lambda-1\right)\left[\left(2A+1\right)k+\left(A+1\right)c\right]}{\left[1-2A+2D+\left(A-2D\right)\lambda\right]k+\left[2D\left(1-\lambda\right)-A\left(2-\lambda\right)\right]c}\,.\label{eq:Y_corr}
\end{equation}
\end{widetext}

Note the explicit dependence on the expansion rate (recall that we are generically assuming $a\propto t^{\lambda}$). The $A$,
$B$, and $D$ parameters come from writing the energy loss terms as 
\begin{equation}
f_{0}+s-f\sim AY\label{eq:A_def}
\end{equation}
\begin{equation}
f_{0}+s+f\sim2(1+BY)\label{eq:B_def}
\end{equation}
\begin{equation}
f_{0}+s\sim1+DY\,.\label{eq:D_def}
\end{equation}
In particular, we can use the physical requirement that $Y$
be positive to impose constraints on the linear term in the expansion
of $s\left(Y\right)$. If we assume Eqns. (\ref{wig_b5gnb}) and (\ref{wig_b5gna}) (meaning $A\sim-1/2+D$, $B\sim1/4+D/2$,
and $DY\sim s$), then 
\begin{widetext}
\begin{equation}
Y\sim\frac{2\left(\lambda-1\right)\left[4kD+\left(1+2D\right)c\right]}{\left(4-\left[1+2D\right]\lambda\right)k+\left(2-\lambda-\left[4\lambda^{2}-6\lambda+4\right]D\right)c}\label{eq:Y_corr_GN}
\end{equation}
\end{widetext}
and now the requirement that both $D$ and $Y$ be positive
implies that the denominator in Eqn. (\ref{eq:Y_corr_GN}) must be negative. If
we further impose that the denominator be large (so that $Y$ may
be small) then 
\begin{equation}
D=\lim_{\mu\rightarrow1^{+}}\frac{s\left(\mu\right)}{\mu-1}\gg\frac{\left(4-\lambda\right)\left(k/c\right)+2-\lambda}{2\lambda\left(k/c\right)+4\lambda^{2}-6\lambda+4}\,.\label{eq:eps_min}
\end{equation}
Note also that 
\begin{equation}
\lim_{D\rightarrow\infty}Y\sim\frac{\left(1-\lambda\right)\left[2\left(k/c\right)+1\right]}{\lambda\left(k/c\right)+\left(2-3\lambda+2\lambda^{2}\right)}\,.\label{eq:lim_Y}
\end{equation}

Recall that as in the standard Goto-Nambu scaling solutions $k$ is itself a function of velocity (and here it may also further depend on the wiggliness) so the above are all implicit solutions. Indeed, in this discussion we have further assumed that $f$, $f_0$ and $s$ are $\mu$-dependent.

If we assume, for example $v\sim0.6$, then these results
yield $D\gg2$, $Y_{rad}\rightarrow1.21$, and $Y_{mat}\rightarrow0.73$
- which means there are surely realistic cosmological scenarios in
which this linear regime approach is not sufficient and a full wiggly
model is necessary. Note, however, that small variations of the scaling
velocity can change these values enough for this linear approach to
suffice (for example, if $v\sim0.7$ then $D\gg0.9$, $Y_{rad}\rightarrow0.58$,
and $Y_{mat}\rightarrow0.43$).

\section{\label{concl}Conclusions}

In this work we have taken the first steps towards modeling the evolution of more realistic cosmic string networks. We have built upon previous work in the literature and described the formalism required for a wiggly extension of the VOS model for Goto-Nambu cosmic strings, which can describe the evolution of small-scale structure on string networks.

While an analysis of the solutions of the full model is left for a companion paper, here we have provided two simpler but physically relevant limits of the geberal formalism. Specifically we have discussed model solutions in the extreme limit where the wiggles make up a high fraction of the total energy of the string network (which physically corresponds to the limit where the tension is arbitrarily small) as well as the opposite (linear) limit where the wiggles only make up a small fraction of the total energy of the network.

As one would expect--aind is confirmed by our analysis of the linear limit of the model---whether or not small-scale structure reaches a scaling regime depends on how much of it is produced by intercommutings, and on how much of it is removed by loop production and gravitational radiation. It is important to realize that there is a further degree of freedom, which is provided by the expansion rate of the universe. Comparing our results for the radiation and matter epochs, and assuming the expected behavior of the phenomenological parameters, this solution would predict larger small-scale structure on correlation length scales in the radiation case. (Indeed, this would be the cas for any scaling length scale, and not just for the scale of the correlation length.) At the qualitative level this in agreement with numerical simulations \cite{BB,AS,FRAC}.

All this being said, we stress that the above analysis is only approximate, and meant to provide some intuition for the phenomenology of the model in the linear limit. The above results are consistent (at least qualitatively) with the results of existing numerical simulations, although these simulations probably have more small-scale structure than can be reliably described by such a linearized model. Thus a more robust comparison, allowing for a detailed test of this linear limit agains simulations, will most likely require custom-run simulations. On the other hand, one should be able to compare exisitng and future simulations to the full wiggly model, which we will discuss in detail in a subsequent paper.

\begin{acknowledgments}
This work was done in the context of project PTDC/FIS/111725/2009 (FCT, Portugal). CJM is also supported by an FCT Research Professorship, contract reference IF/00064/2
012, funded by FCT/MCTES (Portugal) and POPH/FSE (EC). JPV is supported by the Gulbenkian Fundation through {\it Programa de Est\'{\i}mulo \`a Investiga\c c\~ao 2013}, grant number 132590.
\end{acknowledgments}

\bibliography{paper1}

\end{document}